\PassOptionsToPackage{table, dvipsnames, pdftex}{xcolor}
\documentclass[aps,prappl,superscriptaddress, reprint]{revtex4-2}

\usepackage{tikz}
\usepackage{physics}
\usepackage{float}
\usepackage{subcaption}

\usepackage{framed}

\usepackage{xargs}

\usepackage[colorinlistoftodos,prependcaption,textsize=tiny]{todonotes}
\newcommandx{\unsure}[2][1=]{\todo[linecolor=red,backgroundcolor=red!25,bordercolor=red,#1]{#2}}
\newcommandx{\change}[2][1=]{\todo[linecolor=blue,backgroundcolor=blue!25,bordercolor=blue,#1]{#2}}
\newcommandx{\info}[2][1=]{\todo[linecolor=OliveGreen,backgroundcolor=OliveGreen!25,bordercolor=OliveGreen,#1]{#2}}
\newcommandx{\improvement}[2][1=]{\todo[linecolor=Plum,backgroundcolor=Plum!25,bordercolor=Plum,#1]{#2}}
\newcommandx{\thiswillnotshow}[2][1=]{\todo[disable,#1]{#2}}

\begin{document}

\title{Reconfigurable circular polarization medium frequency atomic receiver using magneto-electric effect}

\author{Sujit Garain}
\email[]{sujit.garain@niser.ac.in}
\affiliation{School of Physical Sciences, National Institute of Science Education and Research Bhubaneswar, Jatni-752050, India}
\affiliation{Homi Bhabha National Institute, Training School Complex, Anushaktinagar, Mumbai 400094, India}

\author{Surya Narayan Sahoo}
\email[]{suryans@niser.ac.in}
\affiliation{School of Physical Sciences, National Institute of Science Education and Research Bhubaneswar, Jatni-752050, India}
\affiliation{Homi Bhabha National Institute, Training School Complex, Anushaktinagar, Mumbai 400094, India}
\author{Ashok K Mohapatra}
\email[]{a.mohapatra@niser.ac.in}
\affiliation{School of Physical Sciences, National Institute of Science Education and Research Bhubaneswar, Jatni-752050, India}
\affiliation{Homi Bhabha National Institute, Training School Complex, Anushaktinagar, Mumbai 400094, India}

\date{\today}

\begin{abstract}
Nonlinear magnetoelectric effect(NME) in alkali atomic vapor has applications in precision magnetometry in the radio-frequency domain. We report the application of the NME in alkali atomic vapors for projective measurement of medium-frequency (MF) magnetic fields in a circular basis with an extinction ratio up to 500:1 . Utilizing a longitudinal static magnetic field, we demonstrate a high-sensitivity technique for characterizing the ellipticity of radio-frequency (RF) magnetic fields which can in turn be used for phase sensitive detection in mid frequency communication. Additionally, we demonstrate the conversion of binary phase shift keyed RF magnetic fields into amplitude modulation of generated optical fields, a versatile receiver for communication using the medium frequency band.
\end{abstract}

\maketitle

Quantum sensors based on atomic and optical physics opens new door to technology and precision metrology that could lead fundamental physics \cite{QuantumSensingEssay}. Quantum enhanced electric field sensors have potential impact in communications and remote sensing \cite{RydElecSensing}. Electric fields sensors based on electromagnetically induced spectrum of microwave-dressed Rydberg atoms \cite{AKM_PRL}, have demonstrated excellent sensitivity in the THz and GHz frequency bands \cite{Lin2023, Legaie2024}. Many communication applications require phase detection \cite{Rohde2016} and to enable the same super-heterodyne \cite{Jing2020} and optical-bias schemes \cite{OptBias} have been implemented. The former uses local oscillator microwave directly on the atom reciever while the latter mixes the local oscillator with detected signal of the optical field. The interaction of radio frequency electric fields with atoms is strong due to high transition electric dipole moment which requires high principal quantum number. In MHz regime, existing thermal Rydberg atomic sensors operates in off-resonant regime with limited sensitivity \cite{RydElecSensing, QuantumLimitedAtomicReciever}. Further, the fraction of atoms excited to the Rydberg state is typically low. On the other hand, magnetic dipole interactions only depend on the spherical components and can be used in lower frequency applications to detect the magnetic field component of the radio frequency wave.

Over the past decade, atomic magnetometers have achieved sensitivities rivaling state-of-the-art superconducting quantum interference devices (SQUIDs) \cite{OPMreview_Budker_2007}. This remarkable progress stems from the exquisite sensitivity (femto-tesla level) attainable in a compact form factor, making atomic magnetometers highly attractive for diverse industrial and biomedical applications \cite{OPMreview_Zhao_2023, MEG_Tierney_2019, OPMreview_Romalis_2022, Kitching2008, Aslam2023}. Optically pumped atomic magnetometers (OPMs) exploit the interaction of light with alkali atomic media, where an external magnetic field induces a rotation in the polarization of a probe beam \cite{OpticalMagnetometryBook_Budker_2013}. This rotation serves as a sensitive measure of the applied field and OPMs based on this principle have achieved exceptional sensitivity down to sub-femtotesla level \cite{subfemto_Romalis_2003}. Such performance is demonstrated even in unshielded environments both in DC and RF domain \cite{Seltzer_3axisunshielded_2004, Keder_unshieldedRF_2014, Eddy_Rushton_2022} paving the way for many in-situ applications.

Atomic magnetometers utilizing the nonlinear magnetoelectric effect (NME) \cite{NME_Sushree_2022} offer pico-tesla to femto-tesla sensitivity over a wide radio frequency bandwidth and significantly higher dynamic range due to their detection principle of measuring light amplitude generated in a coherent atomic process. We explore the polarization and spectral properties of the coherently generated light in the presence of a longitudinal DC magnetic field. The corresponding polarization component is given by $\chi^{eemm}_{ijkl} E_j B_k B_l$, i.e., the generated electric field depends on two magnetic fields and one electric field. We demonstrate that by carefully choosing the direction of this longitudinal field, the atomic RF receiver can be made selectively sensitive to either left or right circular polarization of the RF magnetic field. This ability to distinguish between different polarizations presents exciting possibilities for advanced magnetometry applications.

Circularly polarized radio frequency waves in the very high and ultra high frequency band are being routinely used in communication systems. However, the traditional antenna designs for medium frequency or low frequency band would be either hundreds of meters wide or less efficient \cite{Antenna_Balanis_2016, RydElecSensing, Nadeem2021}. The magnetic dipole enabled atomic transitions between Zeeman levels in the same fine structure changes the angular momentum by one unit corresponding to the angular momentum of the RF photon.  We define the quantization axis along $\hat{z}$ and define the left or right circular polarization with respect to it. We show that when the magnetic field is right/left circularly polarized i.e. with relative phase $\pm \pi/2$ between rectilinear components, we obtain its signature in the spectrum of the generated light when we apply a resonant longitudinal magnetic field along $\mp \hat{z}$. Hence, this serve as circular polarization sensor which can be reconfigured to either detect left or right circular polarization.

\begin{figure}[H]
	\centering
	\includegraphics[width=0.99\linewidth]{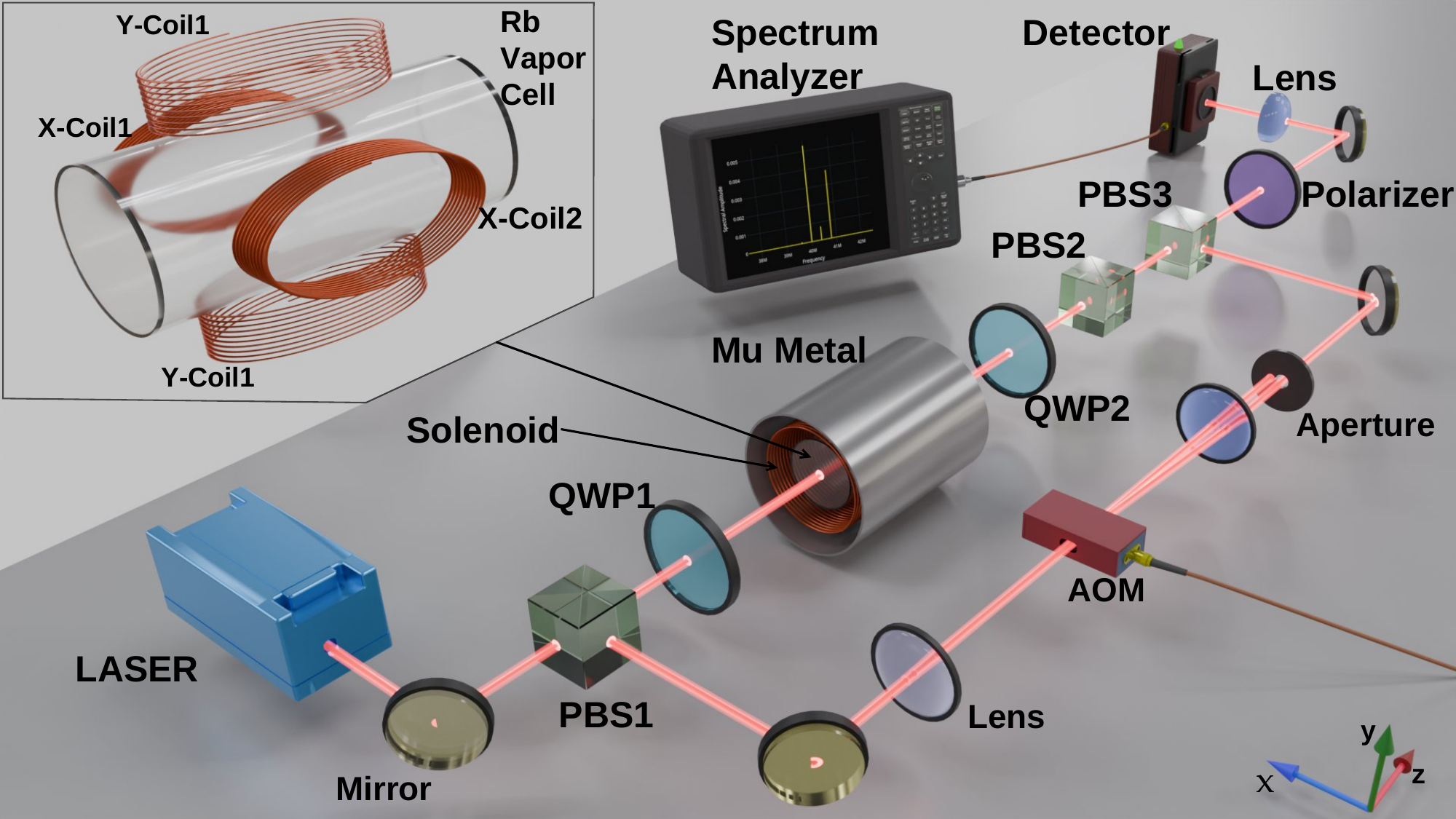}
	\caption{Schematic for the experimental setup for the magnetoelectric effect performed with longitudinal magnetic field. The inset shows the arrangement of the coils around the heated vapour cell to create the polarized RF magnetic fields. The DC magnetic field along $\hat{z}$ is created using the solenoid. The optical left circularly polarized pump beam is filtered to allow the generated light with orthogonal polarization, the magnitude of which is inferred from the spectrum of the resultant beat signal after interference with a local oscillator.}
	\label{fig:fig1}
\end{figure}

The experimental set up for the magneto-electric effect is described in the Figure. \ref{fig:fig1}. We use the Rb 87 D2 transition line from the ground state hyperfine level F=1. The atomic vapor cell (TT-RB-25$\times$75-Q, Triad Technology) was maintained at 85${}^\circ$ C temperature and thus contributions from all the excited state hyperfine levels may not be neglected. However, the effect can be explained by considering only the $F'=0$ state. The laser is locked at wavelength 780.234141 nm using a wavelengthmeter (High Finesse WS-U) within fluctuations of 500 kHz. This frequency is red detuned to the transition $F=1$ to $F'=0$ by $\Delta_p \approx 488 \ \text{MHz}$.
To describe the effect, we choose the quantization axis to be along the direction of propagation of the pump laser beam $\hat{z}$. We prepare the polarization of the optical pump beam to be $\sigma^{(-)}$ or LCP by passing through the polarizing beam splitter PBS1 and the quarter wave plate QWP1 combination. The QWP2 and PBS2 combination is used to ensure that almost all of the pump light is reflected. There was however residual transmission due to the finite extinction ratio  which was characterized to be 250:1. 

The heated vapor cell is placed inside a Mu Metal cylindrical shield which suppresses the ambient DC magnetic field by an order of magnitude. Inside the mu metal shield, a solenoid coil with 80 turns and 8 cm diameter is used to create a spatially uniform longitudinal magnetic field $B_z$. Inside the solenoid, we have a pair of X-Coils and Y-Coils to create radio frequency waves with arbitrary polarization of the magnetic field. The elliptically shaped coils, each with major diameter $8.5 \text{cm}$, minor diameter $3.5 \text{cm}$ and 10 turns creates predominantly fields along $\hat{x}$ and $\hat{y}$ directions. We drive the two pair of X-coils with a function generator (Keysight 33600A) giving a sinusoidal wave at frequency $\nu_{rf}=\omega_{rf}/(2\pi)$. The Y-coils are driven with another frequency coupled channel of the same function generator. The phase $\phi_{rf}$ of the signal driving Y-coil changes the ellipticity of the time varying magnetic field in the $x-y$ plane. The magnetic field generated by these RF coils can be represented as $\vec{B}_{rf} = B_x \cos(\omega_{rf} t) \hat{x} +  B_y \cos(\omega_{rf} t + \phi_{rf}) \hat{y} $. When $B_x$ and $B_y$ are equal, $\phi_{rf}= \pi/2$ (or $3\pi/2$) generates the left (or right) circularly polarized RF magnetic field.

\begin{figure}[H]
	\begin{subfigure}{0.49\linewidth}
		\centering
		\includegraphics[width=0.99\linewidth]{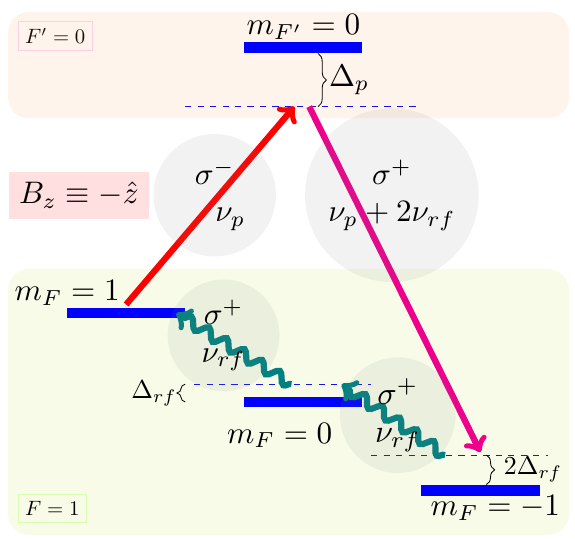}
		\caption{}
		\label{fig:figa}
	\end{subfigure}
	\begin{subfigure}{0.49\linewidth}
		\centering
		\includegraphics[width=0.99\linewidth]{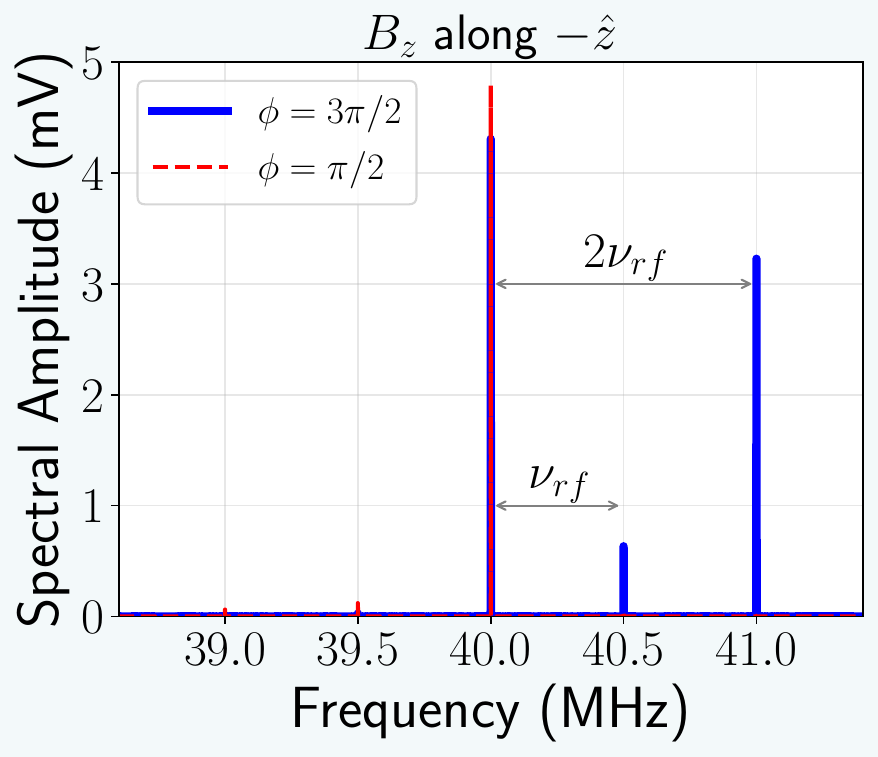}
		\caption{}
		\label{fig:figb}
	\end{subfigure}
	\hfill
	\begin{subfigure}{0.49\linewidth}
		\centering
		\includegraphics[width=0.99\linewidth]{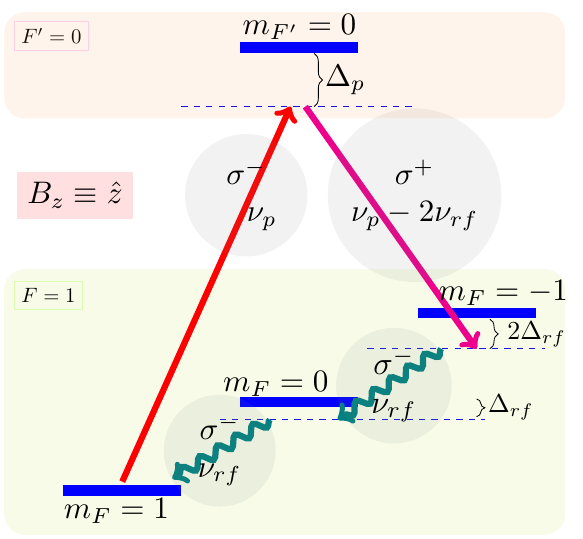}
		\caption{}
		\label{fig:figc}
	\end{subfigure}
	\begin{subfigure}{0.49\linewidth}
		\centering
		\includegraphics[width=0.99\linewidth]{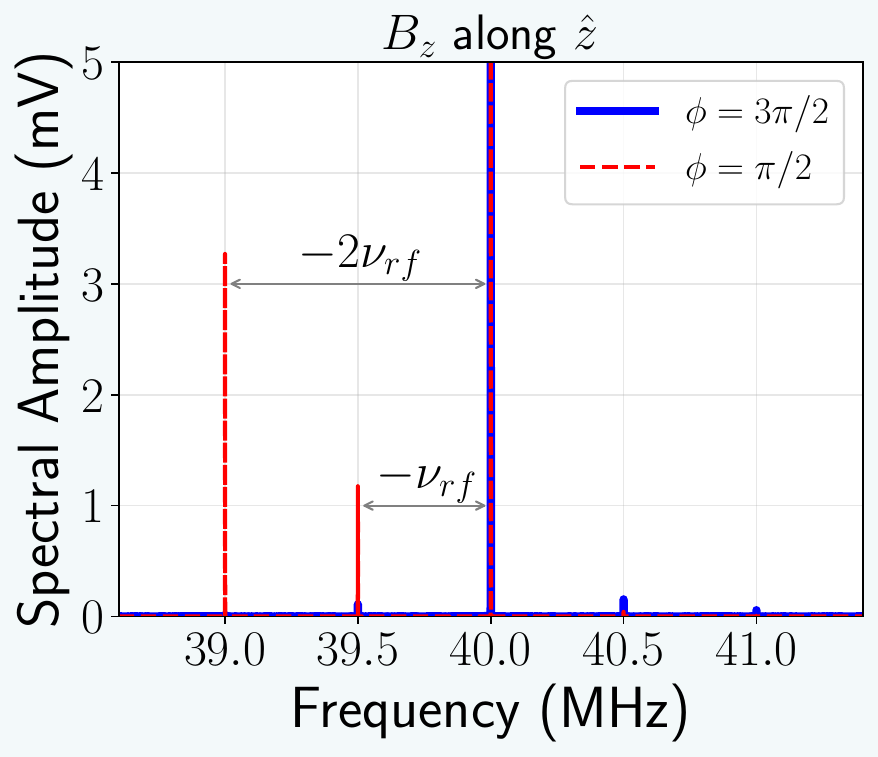}
		\caption{}
		\label{fig:figd}
	\end{subfigure}
	\caption{The energy of ground state Zeeman sublevels, when the DC longitudinal field is along $-\hat{z}$ is shown in (a). The NME effect occurs when the two $\sigma^{(+)}_{rf}$ photons are absorbed. The generated light has higher frequency than the pump and hence the spectral peak appears to its right as observed in (b). When the DC field is  along $\hat{z}$, the $\sigma^{(-)}_{rf}$ photons cause stimulated emission as shown in (c) and the generated light has lower frequency as seen in the spectrum (d). The experiment was performed with $\nu_{rf} = 500 \ \text{kHz}$. The DC magnetic field was tuned so that $\Delta_{rf}$, the detuning of the RF photons from the Zeeman transition resonance is almost zero.}
\end{figure}

In presence of a longitudinal magnetic field $B_z$ created by the current in the solenoid from a low noise constant current source (NGU401, Rhode $\&$ Schwarz), the Zeeman degeneracy of the ground hyperfine state is lifted (See Fig. \ref{fig:figa}). When $B_z$ is negative, the energy level $m_F=1$ increases with the magnitude of the field and that of  $m_F=-1$ decreases. The absorption of photon with RCP magnetic field ($\sigma^{(+)}_{rf}$) can cause the Zeeman transitions from $m_F=-1$ to $m_F=0$ and from $m_F=0$ to $m_F=1$. However, in a thermal atomic system all the Zeeman levels would be almost equally populated. Thus, the $\sigma^{(+)}_{rf}$ photons are equally likely to cause stimulated emission. The $\sigma^{(-)}_p$ polarized beam with frequency $\nu_p=\omega_p/(2\pi)$, excites the atom from the Zeeman state $m_F=1$ in the ground hyperfine level $F=1$ to the excited state $F'=0$. Eventually, due to spontaneous emission the population builds up in the ground state Zeeman levels $m_F=0$ and $m_F=1$. Hence $\sigma^{(+)}_{rf}$ photons gets favorably absorbed due to this population difference created due to optical pumping. Consequently, in presence of both the optical beam and the $\sigma^{(+)}_{rf}$ magnetic field, coherence between the ground states $\ket{F=1, m_F=-1}$, $\ket{F=1, m_F=0}$ and excited state  $\ket{F'=0, m_F=0}$ develops. The atomic polarization due to the coherence between  $\ket{F=1, m_F=0}$ and the excited state may lead to the emission of $\pi$ polarized photons in case the DC magnetic field is not exactly along $\hat{z}$. Therefore, there is little optical emission with frequency $\nu_p + \nu_{rf}$ along $\hat{z}$ axis. On the other hand, coherence between the excited state and the ground state Zeeman level $m_F=-1$ results in emission of $\sigma_{p}^{(+)}$ polarized optical field with frequency $\nu_p + 2 \nu_{rf}$. This emission along $\hat{z}$ is allowed to be transmitted maximally through the QWP2 and PBS2 combination.

When the longitudinal DC magnetic field is positive, optical pumping with $\sigma^{(-)}_p$ polarized beam results in population transfer to the Zeeman state with higher energy. Hence, $\sigma^{(-)}_{rf}$ photons cause stimulated emission from Zeeman level $m_F=-1$ to $m_F=0$ and from $m_F=0$ to $m_F=1$ (See Fig. \ref{fig:figc}). The resultant coherence between the excited state and the ground state $m_F=-1$ leads to emission of  $\sigma^{(+)}_{p}$ polarized optical field with frequency $\nu_p - 2 \nu_{rf}$ along $\hat{z}$. Using heterodyne detection with a local oscillator with frequency $\nu_{LO}$, we are able to spectrally detect the generation of $\nu_p + 2\nu_{rf}$ and/or $\nu_p - 2\nu_{rf}$. Experimentally, the local oscillator is obtained by frequency shifting the pump beam by $40 \ \text{MHz}$ using combinations of acousto-optic modulators. The generated light and the local oscillator which is vertically polarized are combined using PBS3 and then projected to a common polarization using a Polarizer. We use a AC coupled high sensitive avalanche photodiode (APD 210, Menlo Systems) to detect the beat signals. The spectrum of the signal was then observed with the Spectrum Analyzer (Keysight, EXA Signal Analyzer 9010B).

The spectrum when $B_z$ is $-ve$ is shown in Fig. \ref{fig:figb} for the case of $\nu_{rf}=500 \ \text{kHz}$, which shows the peak at $41 \ \text{MHz}$ corresponding to beat signal between the generated light and the local oscillator which is $40 \ \text{MHz}$ away from the probe beam. Note that the generated signal is present when the phase $\phi_{rf} = \pi/2$ and the signal is negligible when the the phase $\phi_{rf} = 3\pi/2$.  There is a small peak at $40.5 \ \text{MHz}$. This appears due to imperfections in the experimental conditions, specifically the presence of small transverse DC fields, which enables the $\pi$ transition component. This process is more efficient as it involves one RF photon compared to the peak shifted by $2\nu_{rf}$ which involves two RF photons. Therefore, small transverse fields can result in significant peaks shifted by $2\nu_{rf}$. The peak at $40 \ \text{MHz}$ corresponds to the beat amplitude due to interference between the local oscillator and the residual pump beam which is not completely removed due to finite extinction ratio of the PBS2 and the slight non-idealness in preparation of the pump polarization. When the magnetic field is along $\hat{z}$ i.e., $B_z$ becomes $+ve$, the residual pump intensity changes due to magneto optical polarization rotations. Therefore, spectral amplitude obtained with this heterodyne technique gives accurate amplitudes of the generated light when compared to the beat signal between the residual pump light and the generated light which appears at frequency $2\nu_{rf}$. With positive $B_z$, we obtain the spectral peak shifted by $-2 \nu_{rf}$ for the phase difference $\phi_{rf} = \pi/2$. This indicates that only $\sigma^{(-)}_{rf}$ photons participate in the process when $B_z$ is $+ve$ and only $\sigma^{(+)}_{rf}$ photons participate when the $B_z$ is $-ve$. Hence, the magneto-electric effect can act as RF receiver, the polarization of which can be reconfigured from LCP to RCP by flipping the direction of the longitudinal magnetic field.

\begin{figure}[H]
	\centering
	\includegraphics[width=0.99\linewidth]{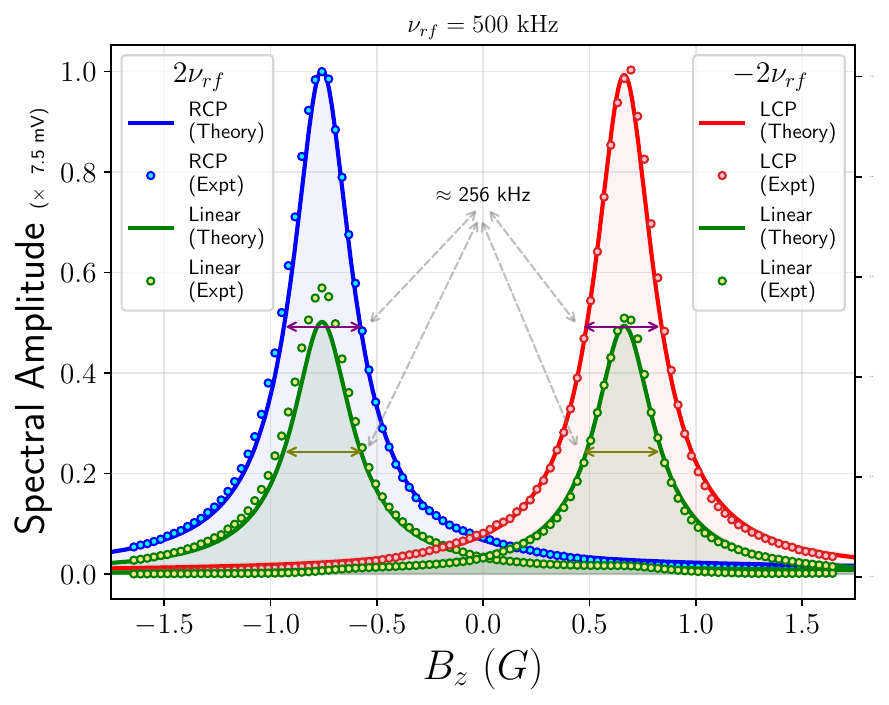}
	\caption{When RF polarization RCP (solid blue line), we obtain the spectral peak shifted by $+2 \nu_{rf}$ when $B_z$ is $-ve$. But the peak is absent for LCP (solid red line). When the DC magnetic field is $+ve$, we have the peak shifted by $-2 \nu_{rf}$ when the magnetic field polarization is RCP (dashed red line). Such a peak is absent for RCP (dashed blue line). In both cases, the extinction ratio was better than 200:1 and sometimes as high as 500:1 (See Supplementary material). When the polarization is linear the $\pm 2 \nu_{rf}$ appears depending on whether $B_z$ is $\mp ve$.}
	\label{fig:fig3}
\end{figure}

The generation of the optical field is efficient when Zeeman energy level spacing due to the longitudinal magnetic field becomes resonant to the frequency of the RF field. In Fig. \ref{fig:fig3}, we plot the spectral amplitude of the peaks shifted by $\pm 2\nu_{rf}$ for the LCP, RCP and LP RF fields. We generate right circularly polarized magnetic fields at $500 \ \text{kHz}$ and as we vary $B_z$ we obtain a lorentzian profile for the spectral peak shifted by $+ 2 \nu_{rf}$. From the fit, we obtain resonance around $-0.765 \ \text{G}$. As we change the magnetic field to $+ve$ side the $- 2 \nu_{rf}$ peak was absent clearing indicating that the direction of the longitudinal magnetic field makes the NME sensitive to RCP RF photons only. Similarly, we obtain the lorentizan peak at magnetic field  $0.670 \ \text{G}$ for the spectral peak shifted by $-2\nu_{rf}$ when the polarization was LCP. No peak was obtained when the polarization was RCP. From the difference in peaks, we could exactly calibrate the residual magnetic field even after placing one layer of the Mu metal shield. The width of the Lorentizan profiles were $0.367 \text{G}$, which corresponds to about $256 \ kHz$.
We define the extinction ratio of this polarization sensitive RF sensor, at any given $\mp B_z$, as the ratio of spectral amplitude at $\pm 2 \nu_{rf}$ peak when the RF is ${}^{R}_{L} CP$. Within the FWHM of the Lorentzian, we obtain the extiction ratio to be more than $200:1$ (and in some regions greater than 500:1) indicating that the polarization sensitivity is robust against small changes in longitudinal field.

To theoretically model the magneto-electric effect, we consider the four-level system comprising of the three Zeeman sub-levels and the excited state \footnote{See Supplementary material for a detailed discission on the theoretical model}.  We use the rotating wave approximation for the interaction of the optical pump beam with the atomic system, but keep the interaction with the RF field as time dependent, which later helps us obtaining solutions where the RF field itself is modulated (See Fig. \ref{fig:fig5}). We use the Lindblad form of the Master equation to solve for the time dependent Hamiltonian using numerical solver diffrax \cite{diffrax}. We used ARC package \cite{alkali_calc} to obtain all the atomic parameters like transition dipole moments and excited state decoherence. The optical and RF field strengths are estimated from the experimental parameters. The ground state decoherence is obtained from the Lorentzian fit of the profiles in Fig. \ref{fig:fig3}. We plot the term of the density matrix corresponding to the coherence states $m_F=-1$ and $m_{F'}=0$ which generates $\sigma^{(-)}$ polarized optical beam and oscillates at $\pm 2\nu_{rf}$  in the rotated frame.

Since the spectral amplitude at resonant positive $B_z$ measures $\sigma^{(-)}_{rf}$ component, we associate it with the projection operator $\Pi_{-}$. Similarly, with negative resonant $B_z$, the measurement is proportional to the $\ev{\Pi_{+}}$. If the RF polarization state is prepared with $B_{x_{rf}} = B_{y_{rf}}$ and some phase $\phi_{rf}$, we can denote the polarization state of the RF photon as $\ket{\psi} =  (1/\sqrt{2})(1,\exp(i \phi_{rf})$. Then the expectation value of the operator $\ev{\sigma_y} = (\ev{\Pi^E_{-}} - \ev{\Pi^E_{+}})/((\ev{\Pi^E_{-}} + \ev{\Pi^E_{+}}))  =  \sin(\phi_{rf})$. Here, $\ev{\Pi^E_{\pm}}$ denotes the experimentally obtained spectral amplitude to due to conversion of $\sigma^{\pm}_{RF}$ photons to the generated light. 

\begin{figure}[H]
	\centering
	\includegraphics[width=0.99\linewidth]{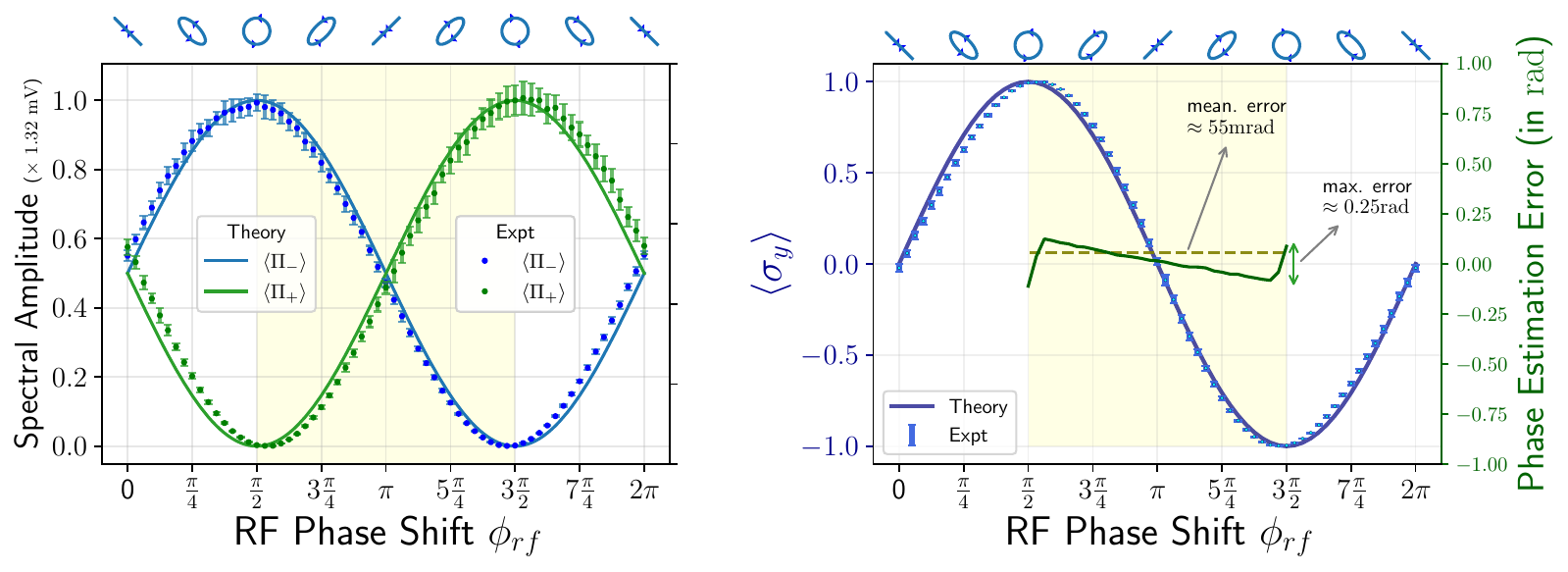}
	\caption{[Left] The spectral amplitude of $+2 \nu_{rf}$ peak with $-ve \ B_z$  is shown in blue as function of the phase difference $\phi_{rf}$ between the sinusoidal currents in X and Y coils. The curve is proportional to $(1+ \sin(\phi_{rf}))/2$. The spectral amplitude of $-2 \nu_{rf}$ peak with $+ve \ B_z$ in green complements the blue curve and is proportional to  $(1- \sin(\phi_{rf}))/2$. The error bars represents statistics over 50 readings acquired over 5 sets of 10 data points each. The solid lines represent the curve obtained from numerical solution to the time-dependent master equation. [Right] To compute $\ev{\sigma_y}$ we take $\ev{\Pi_{+}}$ and $\ev{\Pi_{-}}$ from the same spectrum. Hence the statistical uncertainty is lower. The experimental data closely follows the theoretical prediction for an ideal RF circular projector.  This experiment was performed with $\nu_{rf} = 1 \ \text{MHz}$}
	\label{fig:fig4}
\end{figure}

\begin{figure}[H]
	\begin{subfigure}{0.99\linewidth}
		\caption{}
		\includegraphics[width=0.99\linewidth]{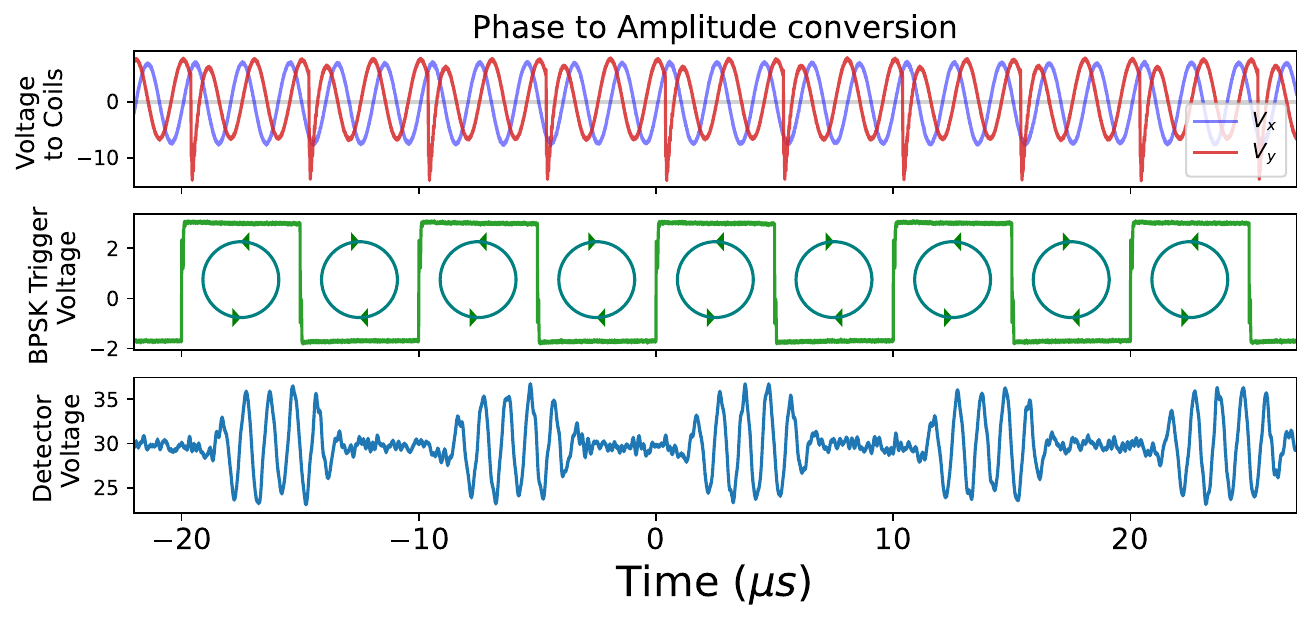}
	\end{subfigure}
	\hfill
	\begin{subfigure}{0.49\linewidth}
		\caption{}
		\includegraphics[width=0.99\linewidth]{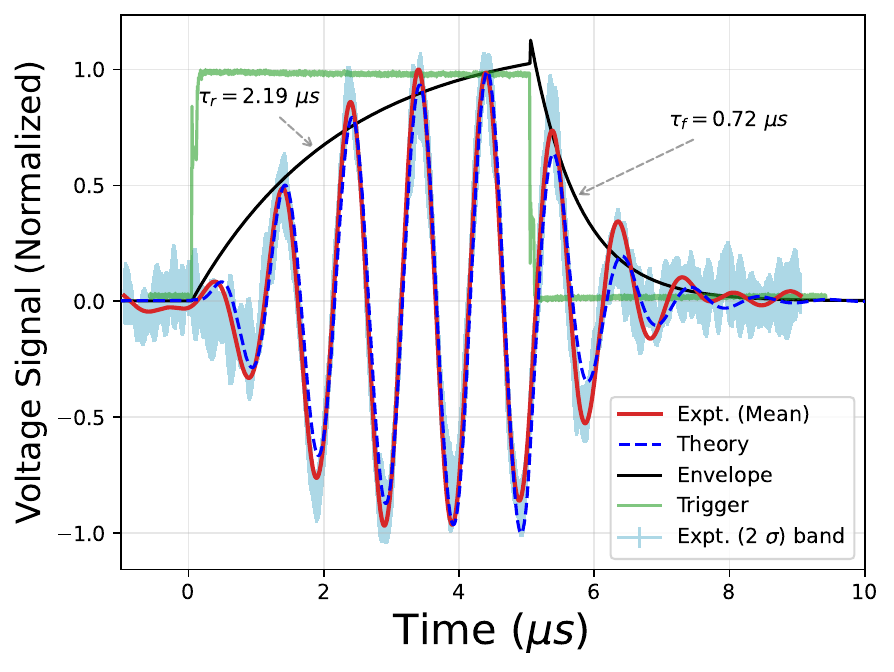}
	\end{subfigure}
	\hfill
	\begin{subfigure}{0.49\linewidth}
		\caption{}
		\includegraphics[width=0.99\linewidth]{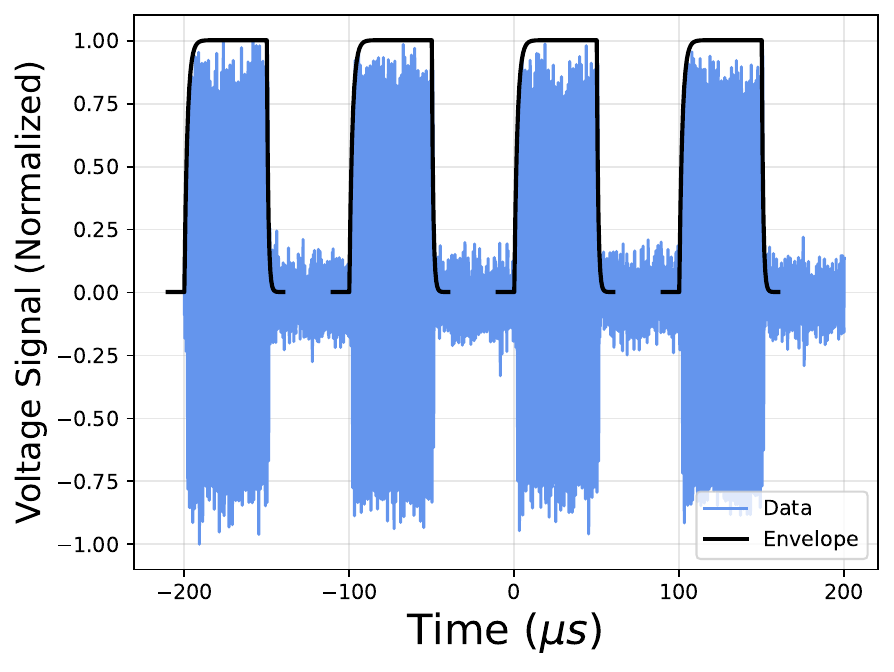}
	\end{subfigure}
	\caption{(a)[Top] The blue curve represents the voltage $V_x$ given to the pair of X-coils with frequency $\nu_{rf} = 500 \ \text{kHz}$. The voltage $V_y$ with the same frequency is given to the Y-coils. The phase of $V_y$ is switched between $90 \deg$ and $270 \deg$. [Middle] The binary phase shift key (BPSK) is controlled by  the trigger shown in green which has a duty cycle of $5 \ \mu s$. When its low, $V_y$ leads $V_x$ by $\pi/2$ resulting in LCP. Similarly, $V_y$ lags $V_x$ by $\pi/2$ when the trigger is high resulting in RCP RF signal. We fix the $B_z$ to resonance for LCP. [Bottom] The detector voltage resulting from beat interference of the generated light with the residual pump. The amplitude of this voltage is high when we have LCP and low when we have RCP. There is slight delay for the BPSK to be effective due to electronics delay. In post-processing, we correct for this delay in (b), where we present the synchronize the trigger and the detected voltage using $V_y$. The detector signal has been rescaled to have unit amplitude. The theory curve represents the time dependent numerical solutions. The black curve denotes the envelope fit to the theory with exponential rise and fall time. In (c), we show the binary phase shift modulation being converted into amplitude modulation with a duty cycle of $50 \mu s$.}
	\label{fig:fig5}
\end{figure}

We can uniquely determine $\phi_{rf}$ in the domain $(\pi/2, 3\pi/2)$ in which the sine function is invertible. We prepare various polarization states by keeping the RF amplitudes in the X coil and Y coil same and only changing $\phi_{rf}$. We took the spectrum once with magnetic field set to resonance on the +ve side and another one with the negative side. The amplitudes of $\pm 2\nu_{rf}$ shifted peaks are shown in Fig . \ref{fig:fig4}. From this,  we compute $\ev{\sigma_y}$ which shows excellent agreement with the prepared phase.

The $\ev{\sigma_y}$ directly corresponds to the ellipticity of the polarization state.  When $\phi_{rf}=\pi/2$, the value is 1 indicating LCP. Similarly, when the value is -1, it corresponds to RCP as obtained in the case when $\phi_{rf}=3\pi/2$. The linear polarization have value 0. Hence, magnetoelectric effect can be used for spectral ellipsometry in the radio frequency domain, where a linearly polarized light can be made incident on a sample and the ellipticity of the reflected field can be used to characterize the sample. Further, if we use one of pair of RF coils as local oscillator, we can infer the phase of the other incoming orthogonally polarized RF magnetic field within $(-\pi/2, \pi/2)$ by measuring $\ev{\sigma_y}$. This technique is similar to superheterodyne measurement \cite{Jing2020} and  opens on wide range of communication applications in the medium frequency band.  Although, as of now the magnetoelectric effect is demonstrated in the medium frequency domain, in principle it can be extended to high frequency microwaves by using Zeeman transitions between ground state hyperfine levels.

Although the first radio communications used medium frequency band, their use has been decreasing due to low data carrying capacity \cite{MidFreq}. But they still are essential for emergency messaging services \cite{Emergency} due to the fact that they are not blocked by structures of the size of buildings. The predominant mode of communication in this frequency band is amplitude modulation which suffers from loss of signal integrity due to reflection  and occlusion from various sources.

 Here, we focus on the process which can be used in communication protocols involving binary phase shift key (BPSK). If the information is encoded in the LCP and RCP which correspond to relative phase shift between the linear components, the generated light amplitude would get modulated accordingly for a fixed $B_z$ polarity. We then observe the beat signal with the residual pump on a fast detector (APD120A, Thorlabs). The amplitude of the signal followed the phase modulation (See Fig. \ref{fig:fig5}). 
With  rectifier followed by appropriate capacitor, we can recover the digital signal after a comparator. Hence, the process results in conversion of phase modulation of the RF field into optical fields which can be converted into amplitude modulation. Further, two such receivers can be placed together - one detecting LCP and the other detecting RCP, which would preserve signal integrity compared to amplitude modulation.

In conclusion, we have demonstrated that using non-linear magnetoelectric effect we can make a RF sensor which is sensitive to the either left or right circular polarization component of the magnetic field. This reconfiguration is achieved by switching the direction of the longitudinal DC magnetic field. The extinction ratio of the sensor is robust against small variations in longitudinal fields making it suitable to be used in presence of stray longitudinal magnetic fields. The setup can be used to measure the ellipticity of the RF polarization and hence can find application in ellipsometry. Using one of the coil as local oscillator, we can also detect the phase of the incident RF field. We also demonstrate that the device can be used to convert phase modulation of one of the linear component of RF, which results in change in circular polarization, can easily be converted into amplitude modulation and thus acting as a digital RF reciever which preserves signal integrity. Compared to traditional mid-frequency antenna receivers, the atomic receiver can be packaged into a portable form factor. This makes them suitable for applications based on magnetic induction tomography \cite{MIT}.

\textit{Acknowledgments} 
SNS would like to acknowledge financial support from the DST-INSPIRE Faculty Fellowship grant No. DST/INSPIRE/04/2023/002297. The authors gratefully acknowledge the financial support from the Department of Atomic Energy, Government of India under the Project Identification No. RIN4001 (National Institute of Science Education and Research Bhubaneswar).

\bibliography{ellipsometryRef.bib}

\end{document}